\documentclass[conference]{IEEEtran}
\usepackage{graphicx}
\usepackage{amsfonts}
\usepackage{amsmath,amsthm}
\usepackage{subeqnarray}
\usepackage{cases}
\usepackage{booktabs}
\usepackage{subfigure}
\ifCLASSINFOpdf
\else
\fi
\begin{document}
\title{Delay-aware data transmission of multi-carrier communications in the presence of renewable energy}

\author{
\IEEEauthorblockN{Tian Zhang}
\IEEEauthorblockA{School of Information Science and Engineering\\
Shandong Normal University\\
Jinan, China 250014\\
Email: tianzhang.ee@gmail.com}
%\and
%\IEEEauthorblockN{Wei Chen}
%\IEEEauthorblockA{Department of Electronic Engineering\\
%Tsinghua University\\
%Beijing, China 100084\\
%Email: wchen@tsinghua.edu.cn}
%\and
%\IEEEauthorblockN{James Kirk\\ and Montgomery Scott}
%\IEEEauthorblockA{Starfleet Academy\\
%San Francisco, California 96678-2391\\
%Fax: (888) 555--1212}
}


%Telephone: (800) 555--1212, Fax: (888) 555--1212}


\maketitle

\begin{abstract}
%\boldmath
In the paper, we investigate the delay-aware data transmission in renewable energy aided multi-carrier system. Besides utilizing the local renewables, the transmitter can also purchase grid power.
By scheduling the amount of transmitted data (The data are stored in a buffer before transmission), the sub-carrier allocation, and the renewable allocation in each transmission period, the transmitter aims to minimize the purchasing cost under a buffer delay constraint.   By theoretical analysis of the formulated stochastic optimization problem, we find that transmit the scheduled data through the subcarrier with best condition is optimal and greedy renewable energy is approximately optimal. Furthermore, based on the theoretical derives and Lyapunov optimization, an on-line algorithm, which does NOT require future information, is proposed. Numerical results illustrate the delay and cost performance of the proposed algorithm. In addition, the comparisons with the delay-optimal policy and cost-optimal policy are carried out.

\end{abstract}


% \fi
\IEEEpeerreviewmaketitle

\section{Introduction}
With the rapid growth of traffic (e.g., speech, data, video, etc.) in wireless communications, the power consumption becomes huge, which has incurred severe environmental problems. Meanwhile, high-rate based new business, such as mobile internet, cloud computing, and big data service, make the increasing trend for power consumption more apparent.
Improving energy efficiency has been an important aim in communication system design\cite{IEEE CommunS&T13: Daquan Feng}. On the other hand, QoS guarantee such as delay is also important especially for delay-sensitive traffic. For example, in real-time multimedia services (e.g., video transmission), if a received packet violates its delay limit then it is considered useless and must be discarded. Generally speaking, power saving and QoS (i.e., delay) improvement are two conflicting aspects in wireless communications: Reducing power consumption will degrade delay performance, and the delay performance improves at the expense of increasing power consumption. Power and delay tradeoff is a fundamental problem in wireless communications\cite{IEEE IT02:R. Berry}\cite{IEEE TC13:X. Zhang and J. Tang}.

Renewable energy has attracted much attention due to its naturality, renewable and pollution-free characteristics. Energy harvesting technique is capable of converting the renewable energy from the environment into electrical energy\cite{IEEE IEEE CommunS&T14:R. Venkatesha Prasad Shruti Devasenapathy Vijay S. Rao and Javad Vazifehdan}.
Recently, incorporating renewable energy in wireless communications system (or energy harvesting aided wireless communications) becomes a hot topic in the literature\cite{IEEE INFOCOM14: H. Li C. Huang S. Cui and J. Zhang}\cite{IEEE TVT14: Tian Zhang}. Meanwhile, renewable energy aided base station begins to leak in practical cellular mobile communications systems.

High data rates communications is significantly limited by inter-symbol interference (ISI) due to the existence of the
multiple paths. Multi-carrier modulation techniques, including orthogonal frequency division multiplexing (OFDM) modulation are considered as the most promising technique to combat this problem. Multi-carrier modulation techniques (e.g.,OFDM) have been widely selected as the physical layer technique in broadband wireless system (e.g., LTE).

The joint investigation of power and delay in multi-carrier communications system has not gained much attention yet, especially when local renewable energy is available. In the paper, we study the delay constrained power allocation in renewable energy aided point-to-point multi-carrier communications. The transmitter is equipped with renewable energy generation device. Meanwhile, the transmitter can purchase power from the grid so as to alleviate the renewable energy's intermittence for a stationary minimum QoS guarantee. The upper layer of the transmitter generates data stochastically, and the data wait in an FIFO (First-In-First-Out) buffer before transmission. In each period, the transmitter decides the number of data for each sub-carrier in the period, and sends to the receiver (Meanwhile, the transmitted data are removed from the buffer). In addition, the transmitter settles how much renewable energy being allocated from the storage battery (and the rest required energy is purchased from the grid). The object is to minimize the cost under a buffer delay constraint. A stochastic optimization problem is formulated accordingly. By theoretical analysis, we prove that transmit all scheduled data through the best subcarrier in each period is optimal. We give the approximate optimal renewable allocation. Furthermore, an on-line algorithm (referred to as BGL algorithm) is proposed based on the theoretical results.

Compared to our previous work \cite{IEEE TVT14: Tian Zhang}, the advances of this paper lie in two aspects:1) Multi-carrier communication is considered, which is vital in ISI alleviation and necessary in frequency selective fading channels. 2)An on-line algorithm combing our theoretical derives and the Lyapunov optimization is proposed.

The rest of the paper is structured as follows: In Section \ref{System model and problem formulation}, the system model is described, and the stochastic problem is formulated accordingly. The analysis of the formulated problem is performed in Section \ref{Problem analysis}. Next, an on-line algorithm, i.e., the BGL algorithm, is proposed in Section \ref{Online algorithm}. Numerical results are given in Section \ref{Numerical results}. Finally, Section \ref{Conclusion} concludes the paper.

%The main symbols utilized in the paper and their meanings are listed in TABLE \ref{Nomenclature}.
% \begin{table*}[]
% \caption{\label{Nomenclature}}
% \centering
% \begin{tabular}{lll}
%  \midrule
%  $M$ & Number of sub-carriers \\
%$A[n]$ & Number of arrived packages during the $n$-th period\\
%$E[n]$ & The amount of generated renewable energy during the $n$-th period\\
%  $\xi[n]$ & Price of the grid power during the $n$-th period \\
%   $R[n]$ & Number of scheduled packages to be transmitted (Sum rate) in the $n$-th period\\
%   $\bold{R}_n=(R_1[n],\cdots,R_M[n])$ & Rate vector over $M$ subcarriers in the $n$-th period\\
%   $W[n]$ & Amount of allocated renewable power \\
%   $Q[n]$ & Length of data buffer at the beginning of the $n$-th period\\
%   $B$ &Capacity of the storage battery\\
%   $E_b[n]$ &Amount of renewable energy in the battery at the beginning of the $n$-th period \\
%    $\mathcal{C}[n]$ &Cost at the $n$-th period\\
%  \bottomrule
% \end{tabular}
%\end{table*}
\section{System model and problem formulation} \label{System model and problem formulation}
\begin{figure}[]
\centering
\includegraphics[width=3.5in]{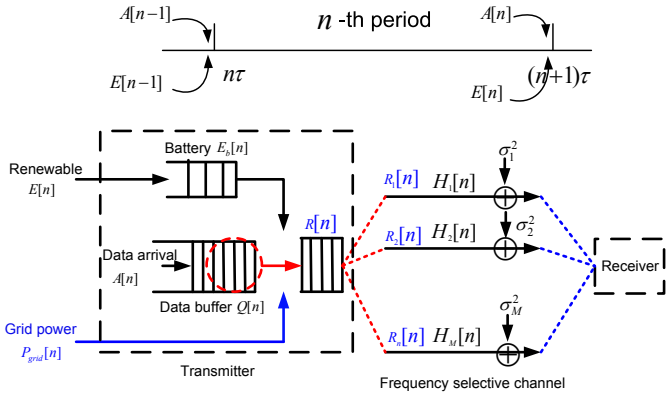}
 %where an .eps filename suffix will be assumed under latex,
% and a .pdf suffix will be assumed for pdflatex; or what has been declared
% via \DeclareGraphicsExtensions.
\caption{Data transmission over multi-carrier system in the presence of renewables}
\label{fig_system model}
\end{figure}
We consider a discrete-time model of point-to-point multi-carrier wireless communications. Time is divided into periods with length $\tau$ each. The $n$-th period is the time interval $\big[n\tau,(n+1)\tau\big)$ as illustrated in Fig \ref{fig_system model}.
There are $M$ sub-carriers, each suffers block flat fading. For a sub-carrier, the channel state remains constant during a period and is variable over periods. The channel power gain of the $i$-th sub-carrier during the $n$-th period is denoted as $H_i[n]$.
$$\bold{H}_n=(H_1[n],\cdots,H_M[n])$$ is the channel state vector of the $M$ sub-carriers during the $n$-th period.
The transmitter is equipped with renewable energy generation device(s), e.g., photovoltaic cells. The renewables are stored in a storage battery with capacity $B$ before usage. The arrived renewables at the end of the $n$-th period is denoted as $E[n]$ ($E[n]$ can be understood as the accumulated renewables during the $n$-th period).
Meanwhile, the transmitter is connected to the power grid, and it can purchase power from the grid. The upper layer of the transmitter generates $A[n]$ packages at the end of the $n$-th period ($A[n]$ can be viewed as the total generated data during the $n$-th period). It is assumed that each package contains $b$ bits. The generated data are stored in an FIFO data buffer. During the $n$-th period, the transmitter removes $R[n]$ packages from the data buffer, and sends to the receiver through the considered $M$ sub-carriers. For the $i$-th sub-carrier, the transmitted data number is $R_i[n]$ during the $n$-th period,\footnote{The rate is $R_i[n]$ package/period/Hz} and $R[n]=\sum_{i=1}^{M}R_i[n]$ is the total scheduled packages for transmission.
$$\bold{R}_n=(R_1[n],\cdots,R_M[n])$$ is the rate vector of the $M$ sub-carriers in the $n$-th period.
Denote the data buffer length at instance $n\tau$ as $Q[n]$. The evolution of the buffer length can be given as
\begin{eqnarray}\label{buffer queue evolution}
Q[n+1]=\big(Q[n]-R[n]\big)^++A[n],
\end{eqnarray}
where $(\cdot)^+=\max\Big\{\cdot,0\Big\}$.
Assume that the additive white Gaussian noise at the $i$-th subcarrier is with zero mean and variance $\sigma_i^2$. The consumed power for reliable transmission (i.e., error-free according to capacity) of the $i$-th sub-carrier is
\begin{eqnarray}
P_i[n]=\frac{\sigma_i^2}{H_i[n]}\big(e^{\theta R_i[n]}-1\big),
\end{eqnarray}
where $\theta=2\ln(2)b/L$ with $L$ being the channel uses in each period.
The total consumed power during the $n$-th period is
\begin{eqnarray}\label{power consumption of data transmission}
P[n]=\sum_{i=1}^{M}P_i[n].
\end{eqnarray}

In the $n$-th transmission, the transmitter allocates $W[n]$ power from the storage battery, and other power, i.e., $\max\Big\{P[n]-W[n],0\Big\}$, is purchased from the power grid. That is to say, the purchased grid power in the $n$-th period is
\begin{eqnarray}
P_{grid}[n]=\big(P[n]-W[n]\big)^+.
\end{eqnarray}
Denote the stored renewables in the battery at instance $n\tau$ as $E_b[n]$. The evolution of the battery energy is
\begin{eqnarray}\label{battery energy queue evolution}
E_b[n+1]=\min\Big\{\big(E_b[n]-W[n]\tau\big)^++E[n],B\Big\}.
\end{eqnarray}
 Denote the grid power price in the $n$-th period as $\xi[n]$, which is constant during the $n$-th period but may change across different periods, then the cost of purchasing the grid power in the $n$-th period is
\begin{eqnarray}
\mathcal{C}[n]=P_{grid}[n]\xi[n]
\end{eqnarray}
The time-average cost over a sufficient large but finite time horizon with $n_{end}$ periods is expressed as
\begin{eqnarray}
\bar{\mathcal{C}} =\frac{1}{n_{end}}\sum\limits_{n =0}^{n_{end}-1}\mathcal{C}[n]
\end{eqnarray}
For notational simplicity, let $Z[n]=\big(\bold{R}_n,W[n]\big)$ be the control action made by the transmitter at the beginning of the $n$-th period. Meanwhile, define $$\mathcal{Z}[n]=\Big\{\big(\bold{R}_n,W[n]\big)\Big|R[n]\le Q[n],W[n]\le\frac{E_b[n]}{\tau}\Big\}$$ as the corresponding action space. $\mu$ is the maximal data buffer length. We have the following constrained stochastic optimization problem.

\begin{eqnarray} \label{optimization problem}
\min_{\left\{Z[n]\right\}_{n=0}^{n_{end}-1}}\bar{\mathcal{C}} =\frac{1}{n_{end}}\sum\limits_{n =0}^{n_{end}-1}\mathcal{C}[n]
\end{eqnarray}
\begin{subequations}
\begin{numcases}{\mbox{s.t.}}
\bar{\mathcal{Q}}=\frac{1}{n_{end}}\sum\limits_{n =0}^{n_{end}-1}
{Q[n]}<\mu,\\
Z[n]\in \mathcal{Z}[n]
\end{numcases}
\end{subequations}

\section{Problem analysis}\label{Problem analysis}

\theoremstyle{lemma} \newtheorem{lemma}{Lemma}
\theoremstyle{theorem} \newtheorem{theorem}{Theorem}

Define a policy as $\pi=(\pi_0,\pi_1,\cdots)$ that $\pi_n$ generates action $Z[n]=\big(\bold{R}_n,W[n]\big)$ in the $n$-th period.
In the problem, a policy includes the rate vector allocation policy, $\bold{R}(\cdot)$ (which generates $\bold{R}_n$ in the $n$-th period), and the renewable allocation policy, $W(\cdot)$(which generates $W[n]$ in the $n$-th period).\footnote{State refers to $\bold{H}_n$, $Q[n]$, $E_b[n]$ and $\xi[n]$; control action is generated according to state under policy} In addition, the rate vector policy can be decomposed into two aspects: How many packages will be transmitted in each period (i.e., the one period rate) and how to allocate these packages over $M$ sub-carriers. They are referred to as the rate allocation policy $R(\cdot)$ (which generates $R[n]$ in the $n$-th period) and the sub-carrier allocation policy, respectively.

Intuitively, to reduce the cost given the total data for the $n$-th period transmission, $R[n]$, all the data should be transmitted through the \lq\lq best\rq\rq~subcarrier for power savings. In addition, power should be allocated from the battery as much as possible since the renewable energy is free. Formally, the one period cost in the $n$-th period is $\Big[\sum_{i=1}^{M}\frac{\sigma_i^2}{H_i[n]}\big(e^{\theta R_i[n]}-1\big)-W[n]\big]^+\xi[n]$. As $R[n]=\sum_{i=1}^{M}R_i[n]$ is fixed, the optimal $\bold{R}_n$ for minimizing $\sum_{i=1}^{M}\frac{\sigma_i^2}{H_i[n]}\big(e^{\theta R_i[n]}-1\big)$ is
$\bold{R}_n^*=\big(R_1^*[n],\cdots,R_M^*[n]\big)$ with
\[{R_j^*}[n] = \left\{ \begin{array}{l}
R[n],j = \mathop {\arg \max }\limits_{i = 1,...,M} \frac{{{H_i}[n]}}{{\sigma _i^2}}\\
0,\rm{otherwise}.
\end{array} \right.\]
Since all scheduled data are transmitted through the subcarrier with best channel condition, we call the strategy as \lq\lq best-subcarrier\rq\rq~ policy.
Apparently, the optimal $W[n]$ for minimizing $\Big[\sum_{i=1}^{M}\frac{\sigma_i^2}{H_i[n]}\big(e^{\theta R_i[n]}-1\big)-W[n]\big]^+$ is allocating the renewable energy as much as possible. And we refer this strategy as greedy renewable energy allocation policy.
Then, we have the following lemma.
\begin{lemma}
Given the transmitting package number in a period, the \lq\lq best-subcarrier\rq\rq~policy and greedy renewable energy allocation policy will be utilized to minimize the one period cost.
\end{lemma}
We have derived that the \lq\lq best-subcarrier\rq\rq~policy and the greedy policy are the optimal sub-carrier allocation policy and optimal renewable energy allocation policy, respectively, for one period given transmitting data number.
Next, there is a natural question \emph{\lq\lq Whether the optimality of the \lq best-subcarrier\rq~\& greedy renewable energy allocation policy holds for the average cost over several periods given the rate allocation policy?\rq\rq}
The following lemma reveals the answer.
\begin{lemma}\label{Lemma:averageoptiforbestgreedy}
Given the rate allocation policy $R(\cdot)$, the \lq\lq best-subcarrier\rq\rq~policy is optimal. In contrast, the greedy renewable policy is NOT always optimal.
\end{lemma}
\begin{IEEEproof}
See the Appendix.
\end{IEEEproof}
\par
\emph{Remark:}
For one period, the \lq\lq best-subcarrier\rq\rq~policy and the greedy policy are optimal sub-policies.
Why the optimality of \lq\lq best-subcarrier\rq\rq~policy can be extended from one to several periods but the greedy renewable policy can not?
The reasons are as follows:
The variation of the subcarrier allocation policy in one period can be thoroughly reflected by the one-period cost, and accordingly does not influence the state of the following period. Then, the optimality for one period can be extended to several periods because of the independence between two consecutive periods.
By contrast, the renewable allocation policy in one period will definitely affect the state (e.g., $E_b$) of the following period, furthermore the corresponding action. That is to say, the variation of the renewable allocation policy in one period can NOT be thoroughly reflected by the one-period cost, the effect can be \lq\lq propagated\rq\rq~ to the following periods. The independence can not hold for the renewable allocation policy. Thus, for greedy renewable policy, optimality in each period does not mean optimality for several periods.

Although the greedy renewable policy is NOT strictly optimal given the rate policy, it is nearly optimal \cite{IEEE TVT14: Tian Zhang}.

Based on the above analysis, we have the following theorem.
\begin{theorem}\label{decomposition of the problem}
The (approximately) optimal solution of (\ref{optimization problem}), $\left\{\bold{R}_n^*,W^*[n]\right\}_{n=0}^{n_{end}-1}$, can be given as follows: Denote $v_n:=\arg\min\limits_{i}\Big\{\frac{\sigma_i^2}{H_i[n]}\Big\}$, $\eta_n:=\frac{\sigma_{v_n}^2}{H_{v_n}[n]}$
and
$P(R[n]):=\eta_n\big(e^{\theta R[n]}-1\big)$.
The optimal rate vector allocation
$\bold{R}_n^*=\big(R_1^*[n],\cdots,R_M^*[n]\big)$ is given by
\[{R_j^*}[n] = \left\{ \begin{array}{l}
R^*[n],j =v_n \\
0,\rm{otherwise},
\end{array} \right.\]
where $R^*[n]$ is the solution of
\begin{eqnarray} \label{optimization problem}
\min_{\left\{R[n]\right\}}\bar{\mathcal{C}} =n_{end}^{-1}\sum\limits_{n =0}^{n_{end}-1}\big[P(R[n])-\min\{E_b[n],P(R[n])\}\big]
\end{eqnarray}
\begin{subequations}
\begin{numcases}{\mbox{s.t.}}
\bar{\mathcal{Q}}<\mu,\\
R[n] \le Q[n].
\end{numcases}
\end{subequations}
The approximately optimal renewable allocation
$W^*[n]=P(R^*[n])-\min\{E_b[n],P(R^*[n])\}.$
\end{theorem}
\begin{IEEEproof}
The theorem can be verified by the first half of Lemma \ref{Lemma:averageoptiforbestgreedy} and the approximate optimality of greedy renewable allocation\cite{IEEE TVT14: Tian Zhang}. The detailed proof is omitted for brevity.
\end{IEEEproof}
\section{On-line algorithm}\label{Online algorithm}

\begin{table}[]
   \caption{}\label{On line algorithm}
   \centering
    \begin{tabular}{lcl}
     \toprule
     \textbf{Algorithm: BGL algorithm} \\
     \midrule
    Step 1: At the beginning of every period $n$, observe the current state including\\
    current environment information (i.e., $\bold{H}_n$ and $\xi[n]$), current data queue length\\
    $Q[n]$, and current battery energy queue length $E_b[n]$.\\
    Step 2: Choose $R[n]\in [0,Q[n]]$ to minimize  \\
 \begin{minipage}{3in}
\begin{eqnarray}\label{Lyp min}
V\cdot\Big[\eta_n\big(e^{\theta R[n]}-1\big)-\tilde{W}[n]\Big]^+\xi[n]-Q[n]R[n],
\end{eqnarray}
\end{minipage}\\
    where $\tilde{W}[n]=\min\bigg\{E_b[n],\eta_n\big(e^{\theta R[n]}-1\big)\bigg\}$, and $V>0$
    is a constant.\\
    Step 3: Denote the selected $R[n]$ in step 2 as $R^*[n]$. Choose the optimal $W[n]$ as \\
    $W^{*}[n]=\min\bigg\{E_b[n],\eta_n\big(e^{\theta R^*[n]}-1\big)\bigg\}$\\
    Step 4:  Update $Q[n]$ and $E_b[n]$ according to (\ref{buffer queue evolution}) and (\ref{battery energy queue evolution}), respectively. \\
     \bottomrule
    \end{tabular}
   \end{table}
In many scenarios, the future information is unavailable. We need to make decisions according to current (and past) information (on-line decision). In this section, we propose an on-line algorithm \lq\lq Best-Greedy-Lyapunov\rq\rq~(BGL algorithm) (in Table \ref{On line algorithm}) based on Theorem \ref{decomposition of the problem} and recently developed Lyapunov optimization \cite{IEEE CDC10: Michael J. Neely}.

The proposed BGL algorithm is purely an on-line algorithm, which requires only the current system state. In the algorithm, we only need to solve a simplified deterministic optimization problem, (\ref{Lyp min}).

First, we analyze the effect of parameter $V$ on the data transmission qualitatively (or semi-quantitatively).
In step 2, $V>0$ is some constant to tradeoff the cost and queue length (i.e., delay). Specifically, given a small value of $V$, queue length is the focus. Then we transmit as much data as possible, $R[n]=Q[n]$. That is to say, we trade cost for delay.
For large $V$, the reducing the cost is dominant, then we do NOT purchase power from the grid and only part (NOT always all) of the buffer data will be transmitted. That is to say, we trade delay performance for cost.

Until now, we have not exactly answer \emph{\lq\lq How much data will be transmitted in each period?\rq\rq} Next, we investigate the problem quantitatively.
That is to say, we concentrate on finding the optimal solution of (\ref{Lyp min}) mathematically to give the answer.

If $$R_{th}:=\frac{1}{\theta}\ln\Big\{\frac{E_b[n]}{\eta_n}+1\Big\}\ge Q[n],$$ (\ref{Lyp min}) is reduced to
\begin{eqnarray}
\min\limits_{0 \le R[n]\le Q[n]}-Q[n]R[n].
\end{eqnarray}
Thus, the optimal solution is $R^{*}[n]=Q[n]$. Otherwise, if $R_{th}< Q[n]$, (\ref{Lyp min}) becomes
\begin{eqnarray}\label{renewable is not enough for transmitting all Sub-1}
\min\limits_{R_{th} \le R[n]\le Q[n]}V\cdot\Big[\eta_n\big(e^{\theta R[n]}-1\big)-E_b[n]\Big]\xi[n]-Q[n]R[n] \nonumber\\
\end{eqnarray}
combined with
\begin{eqnarray}\label{renewable is not enough for transmitting all Sub-2}
\min\limits_{ 0\le R[n]\le R_{th}}-Q[n]R[n].
\end{eqnarray}
The optimal solution of (\ref{renewable is not enough for transmitting all Sub-2}) is $R^*[n]=R_{th}$, and corresponding object value is $\big(-Q[n]R_{th}\big)$; For (\ref{renewable is not enough for transmitting all Sub-1}), if $$R_{s}:=\frac{1}{\theta}\ln\frac{Q[n]}{\theta V \eta_n\xi[n]}\in [R_{th},Q[n]],$$ the optimal solution is $R^*[n]=R_{s}$\footnote{$R_{s}$ is the stationary point of the object function in (\ref{renewable is not enough for transmitting all Sub-1}).}and the corresponding object value is $\frac{Q[n]}{\theta}-V\eta_n\xi[n]-VE_b[n]\xi[n]-\frac{Q[n]}{\theta}\ln\frac{Q[n]}{\theta V \eta_n\xi[n]}<-Q[n]R_{th}$. If $R_{s}<R_{th}$, $R^*[n]=R_{th}$ and the corresponding object value is $\big(-Q[n]R_{th}\big)$. If $R_{s}>Q[n]$, $R^*[n]=Q[n]$ and the corresponding object value is $V\cdot\Big[\eta_n\big(e^{\theta Q[n]}-1\big)-E_b[n]\Big]\xi[n]-Q[n]Q[n]<-Q[n]R_{th}$. In conclusion,
\begin{eqnarray}\label{OptR}
{R^*}[n] = \left\{ \begin{array}{l}
R_{s},R_{s} \in [{R_{th}},Q[n]]\\
{R_{th}},R_{s} < {R_{th}}\\
Q[n],R_{s} > Q[n]
\end{array} \right.
\end{eqnarray}
\par
\emph{Remark:}
$R_{th}$ is the maximal packet number that can be transmitted when using the stored renewable energy only. $R_{th}>Q[n]$ means that the stored renewables can support transmitting all the buffer data. Based on Theorem \ref{decomposition of the problem}, transmitting all the buffer data is optimal in this scenario. That is to say, $R^*[n]=Q[n]$. On the other hand,
when $R_{th}<Q[n]$, the renewables can NOT support emptying the data buffer, i.e., we may need purchasing power from grid in this case: If $R_{s} < {R_{th}}$ (e.g., $V$ is large), cost is the key factor.
We transmit $R_{th}$ packages only using the renewables and no grid power will be purchased. We trade the delay for cost decrease. If $R_{s} > Q[n]$ (e.g., $V$ is small), the delay requirement is sharp. Together with the renewables, we purchase the deficient power from power grid to empty the data buffer. We trade cost for delay performance. The third scenario, $R_{s} \in [{R_{th}},Q[n]]$, both the cost and delay are important, we can NOT trade one totally for the other. Besides the renewables, we buy some grid power for transmitting $R_{s}$ packages (NOT all buffer data).
\section{Numerical results}\label{Numerical results}
In this section, simulations are carried out to illustrate the performance of the BGL algorithm. First, the simulation settings are given.
Next, we show the delay and cost performance of BGL algorithm. Third, we compare the BGL algorithm against two other algorithms (i.e., DOP policy and COP policy) explained later.
\subsection{Simulation setup}
In the simulations, 3 sub-carriers are considered, i.e., $M=3$, and we set $\tau=1$, $b=1$, $N=5$, $\sigma_i^2=1$.
We consider Rayleigh fading in each sub-channel. That is to say, the power gain of each sub-channel is exponentially distributed. It is assumed that the 3 sub-channels have same mean power gain. The data arrival is i.i.d., and the arrived package number in each period chooses $0,10,20,30$ with probability $0.1,0.5,0.3,0.1$, respectively. The renewable energy arrives i.i.d.\cite{IEEE TIT12: O. Ozel and S. Ulukus}, and the arrived renewable energy in each period is $100,300,500,800$ with probability $0.1,0.6,0.2,0.1$, respectively. The grid power price is $0.02$ and $0.05$ with probability $0.3$ and $0.7$, respectively. The performance is averaged over $n_{end}=10^6$ periods. \par
\subsection{Delay and cost performance of BGL algorithm}

\par
\begin{figure}[!t]
\centering
\subfigure[Delay performance]{\includegraphics[width=2.95in]{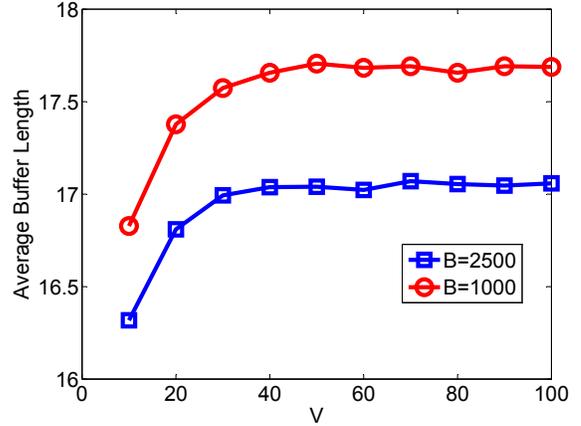}\label{BGLaverageBuff}}
\subfigure[Cost]{\includegraphics[width=2.95in]{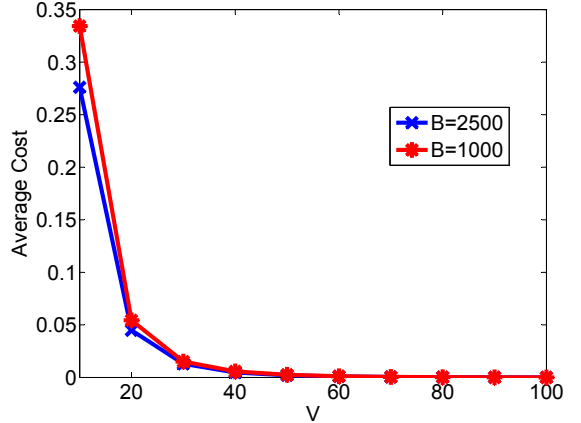}\label{BGLaverageCost}}
\caption{The performance of BGL algorithm versus $V$}
\label{Fig:BGLvsV}
\end{figure}

Fig. \ref{Fig:BGLvsV} plots the average delay and corresponding cost performance of the BGL algorithm with respect to $V$ under different battery capacities.
The mean power gain is set as $0.3$. We consider two battery capacities $B=2500$ and $B=1000$.
We can see that with the increase of $V$, the data buffer length increases at first, and then remains static. Meanwhile, the cost decreases to zero and remains as zero then. It can be explained as follows: $V$ is the trade-off factor. When increase $V$, the \lq\lq importance\rq\rq~of cost increases and the \lq\lq importance\rq\rq~of delay decreases (See (\ref{Lyp min})). The BGL algorithm trades the delay performance degradation for the cost decrease. Mathematically, when $V$ increases, according to (\ref{OptR}), the average transmitted package number decreases at first. Given the mean data arrival and renewable energy arrival, less transmitted data results in longer average data buffer length and less average cost. Hence the data buffer length increases and the cost decreases. When $V$ is very large, $R_{s}<R_{th}$ satisfies almost always, then according to (\ref{OptR}), the transmitted package number remains as $R_{th}$. The average buffer data length remains static given mean data arrival. Furthermore, the renewable energy can support $R_{th}$ and no grid power needs purchasing. Thus, the cost remains as zero. In addition, by comparing the two lines of $B=2500$ and $B=1000$ in each sub-figure, we can find that larger battery capacity improves the delay and cost performance. This is evident since larger capacity stores more (at least no less) renewables and leads to more data transmission and/or less cost.
\par
\begin{figure}[!t]
\centering
\subfigure[Delay performance]{\includegraphics[width=3.0in]{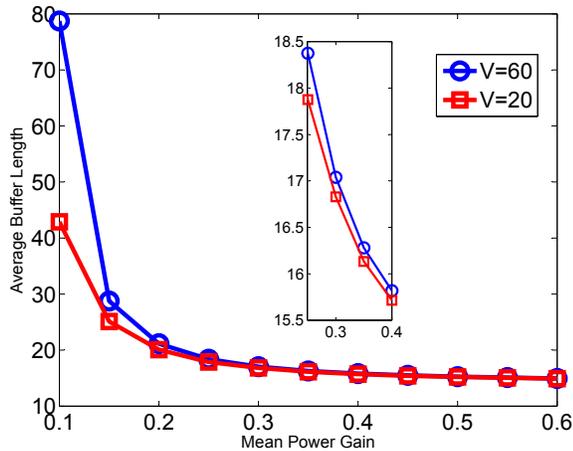}\label{BGLaverageBuff-MPG}}
\subfigure[Cost]{\includegraphics[width=3.0in]{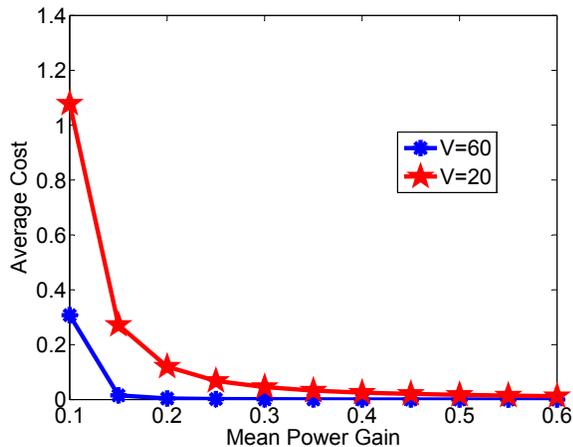}\label{BGLaverageCost-MPG}}
\caption{The performance of BGL algorithm versus mean power gain}
\label{Fig:BGLvsMeanPowerGain}
\end{figure}

Fig. \ref{Fig:BGLvsMeanPowerGain} illustrates the delay and corresponding cost performance of BGL algorithm versus the mean power gain under different $V$. The battery capacity $B=2500$. It is observed that the buffer length and cost decrease when the channel condition improves (i.e., increase of mean power gain) at first, and remains static then. The reason is as follows: When the channel condition improves, the same amount energy supports more data transmission (See (\ref{power consumption of data transmission})). Given the mean data arrival, the average data buffer length decreases. Given the mean renewable arrival, the cost decreases. By comparing the lines of different $V$, we can also find that larger $V$ leads to longer buffer length and less cost. The explanation is same as in Fig. \ref{Fig:BGLvsV}.
\par
\subsection{Algorithm comparison}
In this subsection, we compare the BGL algorithm against two other algorithms described as follows:
\begin{itemize}[]
\item Delay-optimal policy (DOP):
In each period, the transmitter sends all the buffer data through the best sub-carrier, and utilizes the renewable energy as much as possible.
Formally,
\[{R_j}[n] = \left\{ \begin{array}{l}
Q[n],j =v_n \\
0,\rm{otherwise};
\end{array} \right.\]
$W[n]=\min\Big\{E_b[n],\eta_n\big(e^{\theta Q[n]}-1\big)\Big\}$.
\item Cost-optimal policy (COP):
In each period, utilize the renewables only to transmit data as much as possible through the best sub-carrier, and no grid power is purchased. That is to say, the cost is zero. Formally,
 \[{R_j}[n] = \left\{ \begin{array}{l}
\min\{Q[n],R_{th}\},j =v_n \\
0,\rm{otherwise};
\end{array} \right.\]
$W[n]=\min\Big\{E_b[n],\eta_n\big(e^{\theta Q[n]}-1\big)\Big\}$.
\end{itemize}
\par
\begin{figure}[!t]
\centering
\subfigure[Comparison of delay performance]{\includegraphics[width=3.0in]{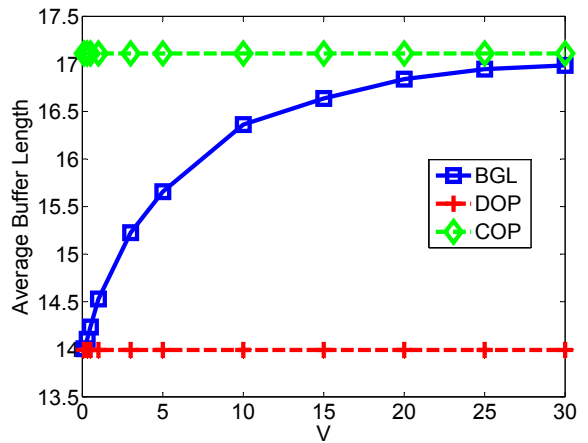}\label{ComBuff}}
\subfigure[Comparison of cost]{\includegraphics[width=3.0in]{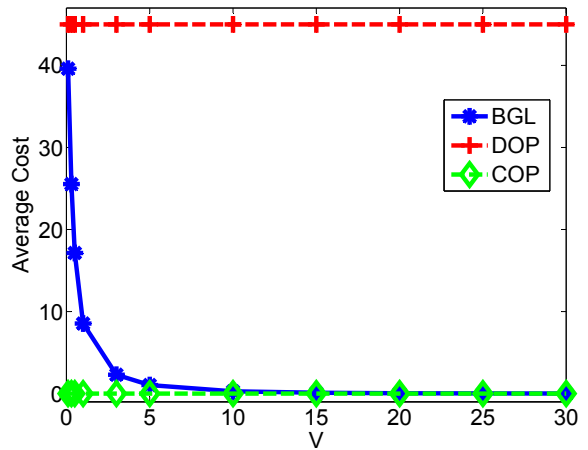}\label{ComCost}}
\caption{The performance comparison against DOP and COP}
\label{Fig:Comparison}
\end{figure}

Fig. \ref{Fig:Comparison} shows the comparisons of delay and corresponding cost, respectively, for BGL algorithm against DOP policy as well as COP policy. The mean power gain is $0.3$, and the battery capacity is $B=2500$. With the increase of $V$, the BGL algorithm reaches the COP policy both in delay and cost. On the contrary, when decrease $V$, the BGL algorithm approaches the DOP policy. In terms of performance, the BGL policy can be viewed as a \lq\lq mixed\rq\rq~ policy of DOP and COP. By adjusting $V$, the delay and cost can be traded off. BGL algorithm with varying $V$ can be applied as follows:
Considering a constraint on delay (i.e., a value of $\mu$), we can get the maximal $V$ from Fig. \ref{ComBuff}. Then we get the optimal cost from \ref{ComCost}.

\section{Conclusion}\label{Conclusion}
Delay-constrained data transmission in multi-carrier communication is studied in the presence of local renewable energy. We formulate a stochastic constrained optimization problem. By theoretical analysis, we derive that transmit data thorough the best sub-carrier is optimal, and greedy renewable allocation is nearly optimal. Additionally utilizing Lyapunov optimization, the BGL algorithm is proposed. Using a tradeoff factor, the cost and delay performance can be adjusted in the BGL algorithm. In the end, simulations show the effectiveness and advance of the BGL algorithm.


\appendix[Proof of Lemma \ref{Lemma:averageoptiforbestgreedy}]
First, the optimality of the \lq\lq best-subcarrier\rq\rq~policy can be verified by contradiction.
Suppose that an optimal policy of the problem (\ref{optimization problem}) does not utilize \lq\lq best-subcarrier\rq\rq~policy as its subcarrier allocation policy. Consider a period, e.g., the $i$-th period, since $R(\cdot)$ is given, $R[i]$ is fixed, and is not transmitted totally through the best subcarrier. Now let a \lq\lq novel\rq\rq~policy that transmit all the $R[i]$ data through the best subcarrier and keep others same as the supposed optimal policy. Then there will be some power savings, and the cost in the $i$-th period will be not more. Meanwhile the delay is the same. As the \lq\lq novel\rq\rq~policy only changes the sub-carrier allocation, it will not change the beginning state of the following periods. In this sense, the periods are independent, i.e., for the following periods, the cost under the \lq\lq novel\rq\rq~policy for each period is not more and delay is the same (Note that the novel policy is feasible) compared to the supposed optimal policy. As $n_{end}$ is sufficiently large, there is at least one period, the cost is less.\footnote{If the transmitter purchases from the power grid in a period under the supposed optimal policy, then in this period, the \lq\lq novel\rq\rq~policy produces less cost.} Hence the average cost is less (Observe that $n_{end}$ is finite). This contradicts with the assumption. The proof of the optimality of the \lq\lq best-subcarrier\rq\rq~policy completes.
\par
Second, the Lagrange relaxed problem of (\ref{optimization problem}) (with multiplier $\lambda>0$) can be expressed as
\begin{eqnarray} \label{Lagrange relaxed problem}
\min_{\left\{Z[n]\right\}_{n=0}^{n_{end}-1},Z[n]\in \mathcal{Z}[n]}\bar{\mathcal{C}} =\frac{1}{n_{end}}\sum\limits_{n =0}^{n_{end}-1}\mathcal{C}[n]+\lambda Q[n]
\end{eqnarray}
The optimal renewable energy allocation policy of (\ref{optimization problem}) must be optimal for (\ref{Lagrange relaxed problem}). In other word, if we can prove that the greedy renewable policy is not optimal for (\ref{Lagrange relaxed problem}), then we prove that the greedy renewable policy is not optimal for (\ref{optimization problem}). The non-optimality of the greedy renewable policy for (\ref{Lagrange relaxed problem}) can be verified similarly as the proof of Lemma 10 in \cite{IEEE TVT14: Tian Zhang}. Then the proof of the greedy renewable policy's non-optimality completes.

%\section*{Acknowledgment}



%\bibliographystyle{IEEEtran}

\end{document}